\documentclass[twocolumn]{svjour3}
\journalname{EPJ}

\usepackage[latin1]{inputenc}
\usepackage{hyperref}
\usepackage{amsmath}
\usepackage{amsfonts}
\usepackage{amssymb}
\usepackage{graphicx}
\usepackage[T1]{fontenc}
\usepackage{siunitx}
\usepackage{doi}

\newcommand*{\affaddr}[1]{#1} 
\newcommand*{\affmark}[1][*]{\textsuperscript{#1}}

\begin{document}
	
\title{Time-of-flight spectroscopy of ultracold neutrons at the PSI UCN source}

\author{G.~Bison\affmark[1], W.~Chen\affmark[1,2], P.-J.~Chiu\affmark[1,2,*], M.~Daum\affmark[1], C.~B.~Doorenbos\affmark[1,2], K.~Kirch\affmark[1,2], V.~Kletzl\affmark[1,2], B.~Lauss\affmark[1], D.~Pais\affmark[1,2], I.~Rien\"acker\affmark[1,a], P.~Schmidt-Wellenburg\affmark[1], G.~Zsigmond\affmark[1,b]\\
{\normalfont\affaddr{\affmark[1]Paul Scherrer Institut, CH-5232 Villigen-PSI, Switzerland}\\
\affaddr{\affmark[2]ETH Z\"urich, CH-8092 Z\"urich, Switzerland}\\
\affaddr{\affmark[*]current address: Universit\"at Z\"urich, CH-8057 Z\"urich, Switzerland}\\}}

\authorrunning{G.~Bison et. al.}

\institute{
	\affaddr{\affmark[a]corresponding author: ingo.rienaecker@psi.ch} \\
	\affaddr{\affmark[b]corresponding author: geza.zsigmond@psi.ch}
}

\date{Received: date / Accepted: date}

\maketitle

\begin{abstract}
	The ultracold neutron (UCN) source at the Paul Scherrer Institute (PSI) provides high intensities of storable neutrons for fundamental physics experiments. The neutron velocity spectrum parallel to the beamline axis was determined by time-of-flight spectroscopy using a neutron chopper. In particular, the temporal evolution of the spectrum during neutron production and UCN storage in the source storage volume was investigated and compared to Monte Carlo simulation results. A softening of the measured spectrum from a mean velocity of 7.7(1)\,\si{\m \per \s} to 5.1(1)\,\si{\m \per \s} occurred within the first 30\,\si{\s} after the proton beam pulse had impinged on the spallation target. A spectral hardening was observed over longer time scales of one measurement day, consistent with the effect of surface degradation of the solid deuterium moderator.
\end{abstract}

\section{Introduction}
\label{sec:introduction}

Neutrons are defined to be ultracold if they can be reflected under all angles of incidence from suitable material surfaces \cite{Zeldovich1959,Lushchikov1969,Steyerl1969}. Such materials have neutron optical potentials of a few hundred neV, corresponding to critical velocities of a few \si{\m \per \s}. The total reflection from surfaces allows the confinement of ultracold neutrons (UCNs) in storage bottles for hundreds of seconds, a technique used for the measurement of fundamental properties of the neutron \cite{Ignatovich1990,Golub1991,Steyerl2020}. These experiments are often limited by statistics, prompting a global effort to enhance the output of existing ultracold neutron sources \cite{Kahlenberg2017,Ito2018,Bison2022a} or utilize new and improved sources currently under construction \cite{Korobkina2014,Ahmed2019,Chanel2023,Frei2023}. A soft energy spectrum and a high number of storable neutrons is crucial to achieve high UCN densities and long observation times in storage experiments.\\

The Paul Scherrer Institute (PSI) operates a spal\-lation-driven, solid deuterium-based source for UCN \cite{Bison2020,Lauss2021}. The source hosts worldwide leading experiments to measure the neutron electric dipole moment \cite{Abel2020}, as well as to search for dark matter candidates, such as axion-like particles \cite{Afach2015,Abel2017a} and oscillations of neutrons into sterile states \cite{Abel2021,Ayres2022}. The thermal moderation \cite{Becker2015}, UCN production \cite{Bison2022a}, as well as UCN storage and transport \cite{Bison2022} in the source were characterized previously. In this paper we report on the measurement of the distribution of the longitudinal velocity component $v = L/T$ of UCNs by time-of-flight $T$ spectroscopy along a flight path $L$ behind a neutron chopper mounted at beamport \text{West-1} \cite{Bison2020}. The results from this measurement are compared to spectra obtained using the UCN source simulation model. Benchmarked simulation models of the source are a valuable input for future experiments. For example, parameters obtained from simulation models like the center-of-mass offset (necessary to estimate systematic effects in \cite{Ayres2021n2edm}) or the mean free flight time (a key parameter to compute the statistical sensitivity for the experiment described in \cite{Ayres2022}) of UCN in a storage volume depend on the correct implementation of the initial velocity spectrum. \\

\begin{figure*}[t]
	\begin{center}
		\resizebox{\textwidth}{!}{\includegraphics{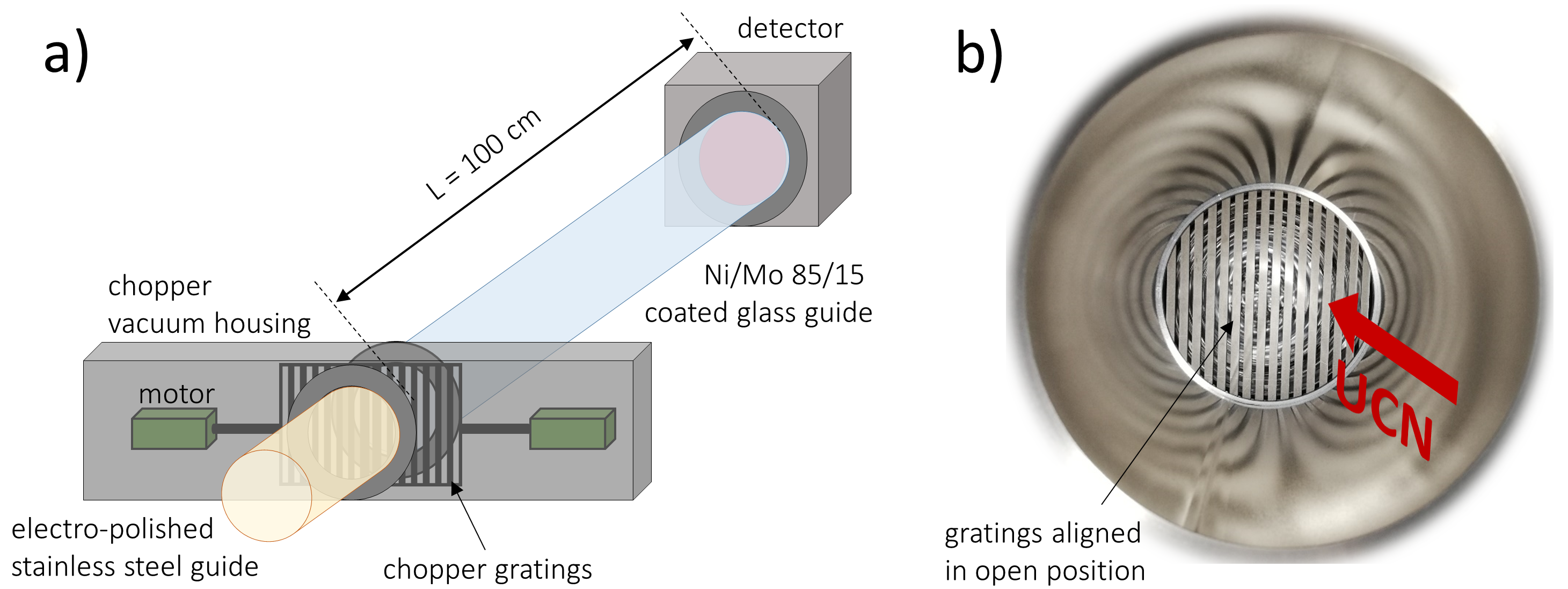}
		}
	\end{center}
	\caption{\textbf{a)} Drawing of the measurement setup, starting from the electro-polished stainless steel guide that was attached to the beamport shutter, up to the CASCADE$^1$ detector. \textbf{b)} A picture of the chopper gratings aligned in the open position, seen from a perspective looking through the short steel guide in front of the chopper towards the detector in the back (behind the chopper). The picture was taken while the whole setup was removed from the beamport shutter.}
	\label{fig:setup}
\end{figure*}

In section \ref{sec:measurement} we present the concept and parameters of the measurement setup, consisting of a UCN detection system and a neutron chopper. We describe the time-of-flight (TOF) data analysis including our background subtraction technique in section \ref{sec:analysis}. In section \ref{sec:results}, we discuss the deduced velocity spectrum and its evolution in time during UCN storage in the UCN source volume \cite{Bison2020}. The influence of UCN source operational procedures on the spectrum is investigated.


\section{Measurement}
\label{sec:measurement}

\begin{figure}[b]
	\begin{center}
		\resizebox{0.45\textwidth}{!}{\includegraphics{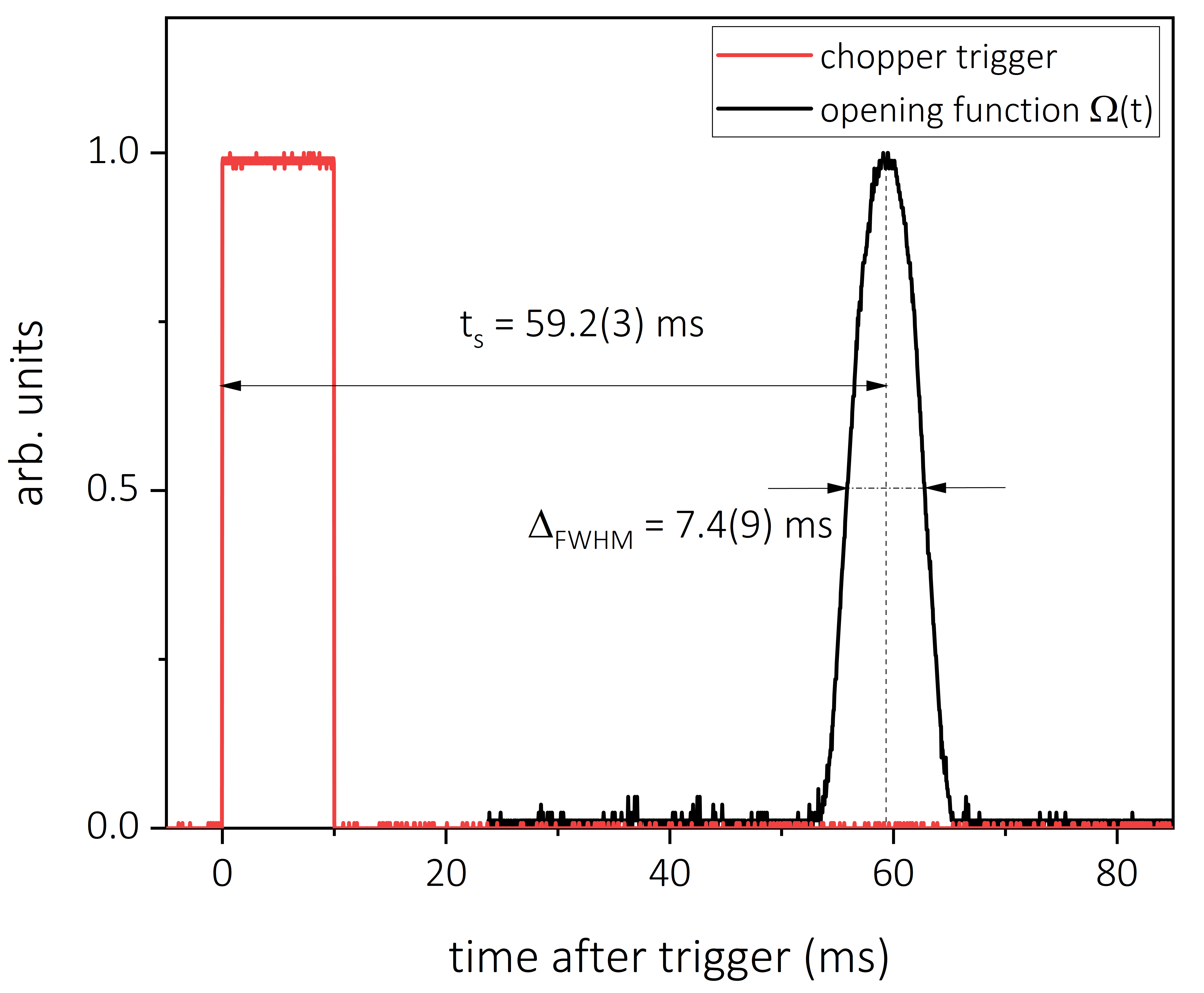}
		}
	\end{center}
	\caption{Plot of the chopper trigger and the subsequent photo diode signal to measure the opening function $\Omega(t)$ of the chopper.}
	\label{fig:opening}
\end{figure}

The measurement setup, consisting of a neutron chopper, a $L=100\,\si{\centi \m}$ flight path, and a neutron detector\footnotemark[1]
is depicted in Fig.\,\ref{fig:setup} a). The chopper is based on the design described in \cite{Lauer2010} and uses the opposing linear motion of two titanium gratings to achieve short opening times. Each grating has 25, $s=3.0\,\si{\milli \m}$ wide, vertical slits, separated by 5.1\,\si{\milli \m}, as shown in Fig.\,\ref{fig:setup} b). The thickness of the gratings is $d=1.3\,\si{\milli \m}$. The first grating is moved by a piston from the left, connected to a linear motor in the vacuum housing of the chopper. The second grating is connected to a second motor on the right and slides directly behind the first grating with a minimal gap of less than 0.4\,\si{\milli \m} between the gratings. The chopper time resolution is determined by its opening function $\Omega(t)$, i.e. the fraction of the cross section of the neutron guide that is unblocked by the gratings as a function of time. When aligned, the 3\,\si{\milli \m} wide slits of the two gratings open a maximum of approximately $\Omega(0)=38~\%$ of the cross section of the guide. The acceleration of each linear motor was set to $a =100\,\si{\m \per \square \s}$ to achieve a full width at half maximum opening time of $\Delta\textsubscript{FWHM} = 2\sqrt{s/(2a)} = 7.7\,\si{\milli \s}$. The opening function is described by a stepwise accelerating motion ($+a,-a,-a,+a$) in four time intervals in which the individual gratings move identical distances of $\frac{s}{4} = \frac{1}{2} a t^2$ each. The FWHM of the entire motion is then given by twice the time per interval. This is consistent with a measurement of the opening time (see Fig.\,\ref{fig:opening}) performed with a photo diode located behind the gratings and a diffuse light source in front. The standard deviation of the opening function was $\sigma_t = 3.5\,\si{\milli \s}$. The time offset, $t_s = 59.2(3)\,\si{\milli \s}$,  between the maximum of a fit to the opening function and the chopper trigger, an electronic signal transmitted by the chopper controller shortly before each opening operation,  was also determined by this method. Additionally, a calibration measurement \cite{Rienacker2022} of the time offset was performed by measuring the UCN count rates, $N_1(v)$ and $N_2(v)$, with two different flight path lengths, $L_1 = 100\,\si{\centi \m}$, $L_2 = 200\,\si{\centi \m}$, using two identical guides of 100\,\si{\centi \m} length. It was found that $\frac{dN_1}{dv}(\frac{L_1}{t_1-t_s}) = \frac{dN_2}{dv}(\frac{L_2}{t_2-t_s})$ at the respective maxima of the normalized velocity distributions, confirming the measured time offset within uncertainties.\\

The chopper was connected to the \text{West-1} beamport of the PSI UCN source with a 20\,\si{\centi \m} long electro-polished stainless steel guide with an inner diameter of 135\,\si{\milli \m}. The guide was connected to the beamport shutter via a flat stainless steel flange, reducing the beamline diameter of 180\,\si{\milli \m} to 135\,\si{\milli \m} by a concentric opening in its center. Similar flanges were used to connect the stainless steel guide to the chopper on one side, and a $L=100\,\si{\centi \m}$ glass guide with inner diameter of $130\,$mm on the other side between chopper and detector, see Fig.\,\ref{fig:setup} a). The glass guide was sputter-coated with Nickel/Molybdenum 85/15 \cite{Bison2020} to obtain a high neutron optical potential. A CASCADE 2D U-200 UCN detector\footnotemark[1] was attached at the end of the 100\,\si{\centi \m} glass guide. The time resolution of the detector was set to $dt = 1\,\si{\milli \s}$. The energy acceptance of the detector has a lower cut-off $E\textsubscript{min} = 54\,\si{\nano \eV}$ due to its 100\,\si{\micro \m} thick AlMg3 entrance window \cite{Bison2020}, corresponding to a lower bound on the velocity component orthogonal to the window of $v\textsubscript{min} = 3.2\,\si{\m \per \s}$.\\

\footnotetext[1]{www.n-cdt.com/cascade-2d-200}

\begin{figure}[b]
	\begin{center}
		\hspace{-0.5cm}
		\resizebox{0.5\textwidth}{!}{\includegraphics{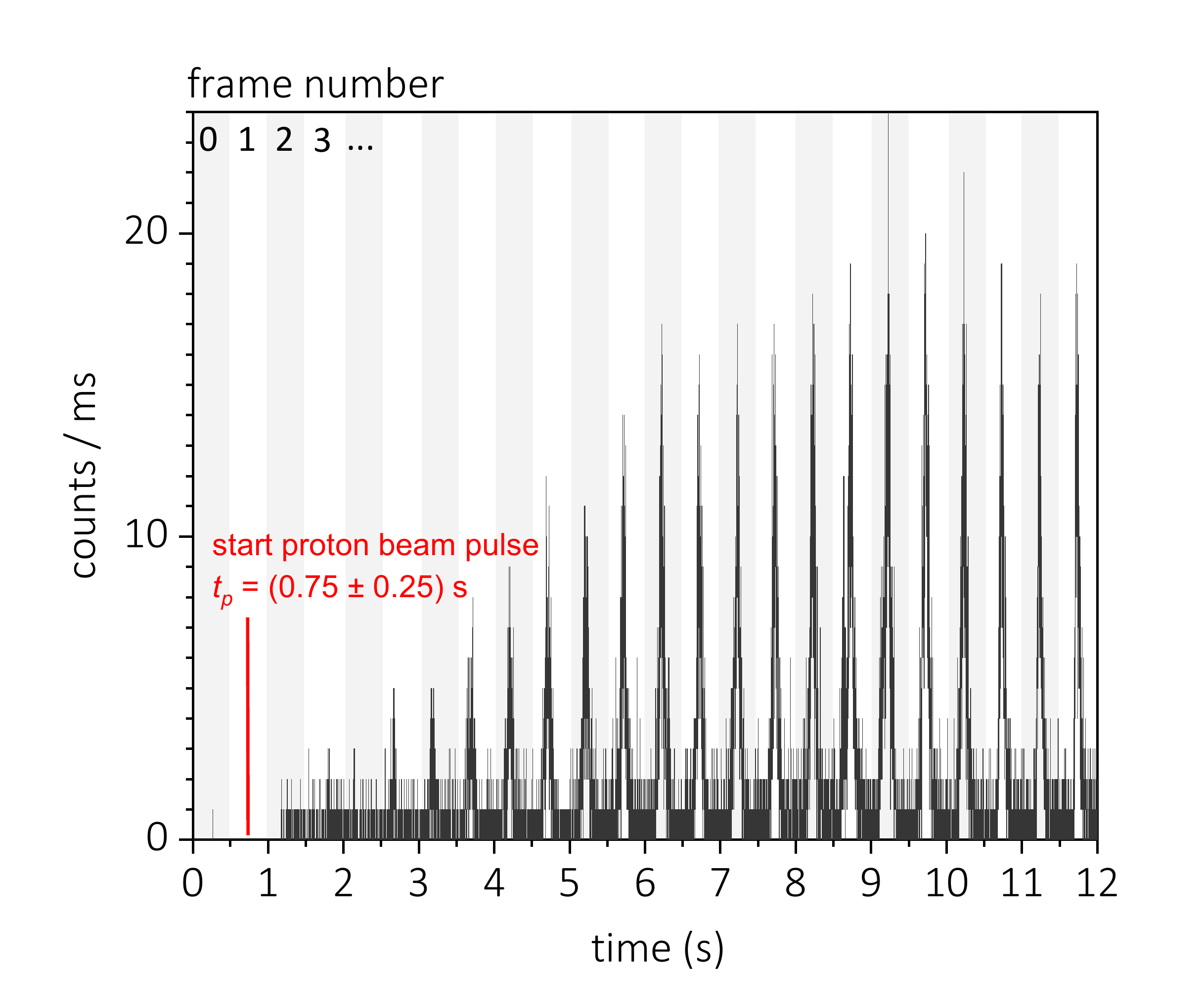}
		}
	\end{center}
	\caption{Raw UCN time spectrum recorded with the CASCADE$^1$ detector during the first 12\,\si{\s} of one cycle with the chopper operating at 2\,\si{\Hz}. On the time axis, $t=0\,\si{\s}$ coincides with the first chopper trigger signal. Relative to that, the 8\,\si{\s} long proton beam pulse starts at $t_p = (0.75 \pm 0.25)\,\si{\s}$, as indicated in the plot. Each cycle is divided into 600 frames, i.e. time intervals $[k\,t_f,(k+1)\,t_f]$ with $k=0,1,...,599$ which contain TOF spectra with respect to the time of the corresponding chopper openings. The UCN count rate increases during the proton beam pulse and slowly decreases afterwards. }
	\label{fig:frames}
\end{figure}

During our spectroscopy measurements we recorded time-of-flight data during approximately one day of uninterrupted standard operation of the PSI UCN source \cite{Bison2020,Lauss2021}. A new measurement cycle was started after each proton beam pulse onto the source's spallation target. The average proton beam current during a pulse was approximately 2.0\,\si{\milli \ampere} and the duration was set to 8\,\si{\s}. The internal clock of the UCN detector was synchronized to the chopper by receiving a coincidence signal between the trigger sent by the chopper and a signal from the proton beam control that rises 1\,\si{\s} before the start of the proton beam pulse and stays active during the pulse. The chopper was running continuously with a duty cycle of $1.5\,$\%, opening the neutron guide at a frequency of approximately 2\,\si{\Hz}. The exact time between two consecutive chopper triggers was measured to be $t_f=499.855(5)\,\si{\milli \s}$, where the uncertainty denotes the observed stability of the signal and the accuracy of a measurement over multiple days. Due to the coincidence with the chopper trigger the start of the measurement may be delayed with respect to the proton beam signal by up to $t_f \approx 500 \, \si{\milli \s}$. Since the chopper was not synchronized to the proton beam operation, this delay may not be constant and we add it as a systematic uncertainty on the start time of the proton beam pulse, i.e. $t_p=(0.75\pm0.25)\,\si{\s}$ as indicated in Fig.~\ref{fig:frames}. \\ 

Several measurements were performed to determine the leakage rate of UCNs through small gaps between the chopper housing and the neutron guides or the gratings in the closed position. Taking the ratio of the count rate measured by the detector with closed chopper versus (permanently) open chopper, one finds an average UCN leakage of $(4\pm1)\cdot 10^{-3}$. The UCN leakage, and other effects like electronic noise and secondary radiation during the proton beam pulse, cause background in the measured TOF spectra, which was fitted and subtracted as discussed in section \ref{sec:analysis}.\\

Since the total thickness of the gratings is on the same order than the horizontal extension of the openings, the chopper collimates the incoming neutron flux. To estimate the effect we have implemented the geometry of the chopper gratings, including several states of partially open gratings according to the measured opening function into our simulation model, which is discussed in more detail in section \ref{sec:results}. We found that the angular cut-off imposed on the transversal velocity components will also affect the observable axial velocity distribution. In particular, UCN with low axial velocities will be suppressed, a finding that is qualitatively consistent with studies and observations by the Nuclear \& Particle Physics group at the Institut Laue-Langevin \cite{Jenke2023} during measurements with a chopper of the same type.

\section{Analysis}
\label{sec:analysis}

The measured count rate in each time bin $dt$ at time $t$ was averaged over all 286 measurement cycles, recorded in a period of approximately one day, to obtain the count rate averages $\frac{dN}{dt}(t)$. The data was divided into 600 individual frames. The TOF spectrum of frame $k$ represents the spectrum at time $t_k = k \, t_f-t_p$ after the start of the proton beam pulse. Each frame contains one TOF spectrum 
\begin{equation}
\frac{dN}{dt}\Big|_{t_k}(T), \quad T = \text{mod}(t,t_f)-t_s \in[-59,441]\,\si{\milli \s},
\label{eq:Spectrum}
\end{equation}
where the TOF axis was shifted by the chopper time offset $t_s\approx59\,\si{\milli \s}$. The division of the data into individual frames is illustrated for the first 24 frames in Fig.~\ref{fig:frames}. The timing uncertainties of $\delta t_f=0.005\,\si{\milli \s}$ and $\delta t_s = 0.3\,\si{\milli \s}$ of the chopper trigger signal discussed in section \ref{sec:measurement} lead to a small contribution to the uncertainty of $\delta T = \sigma_t + k \, \delta t_f + \delta t_s $ on the time of flight in frame $k$. We have used the standard deviation $\sigma_t$ of the opening function as an estimate of the timing uncertainty for the case that no deconvolution with the chopper opening function is applied. The associated uncertainty on the velocity bins of the deduced velocity spectrum is then given by $\frac{L\delta T}{T^2}$ and evaluated in section\,\ref{sec:results}.\\

\begin{figure}[t]
	\begin{center}
		\resizebox{0.48\textwidth}{!}{\includegraphics{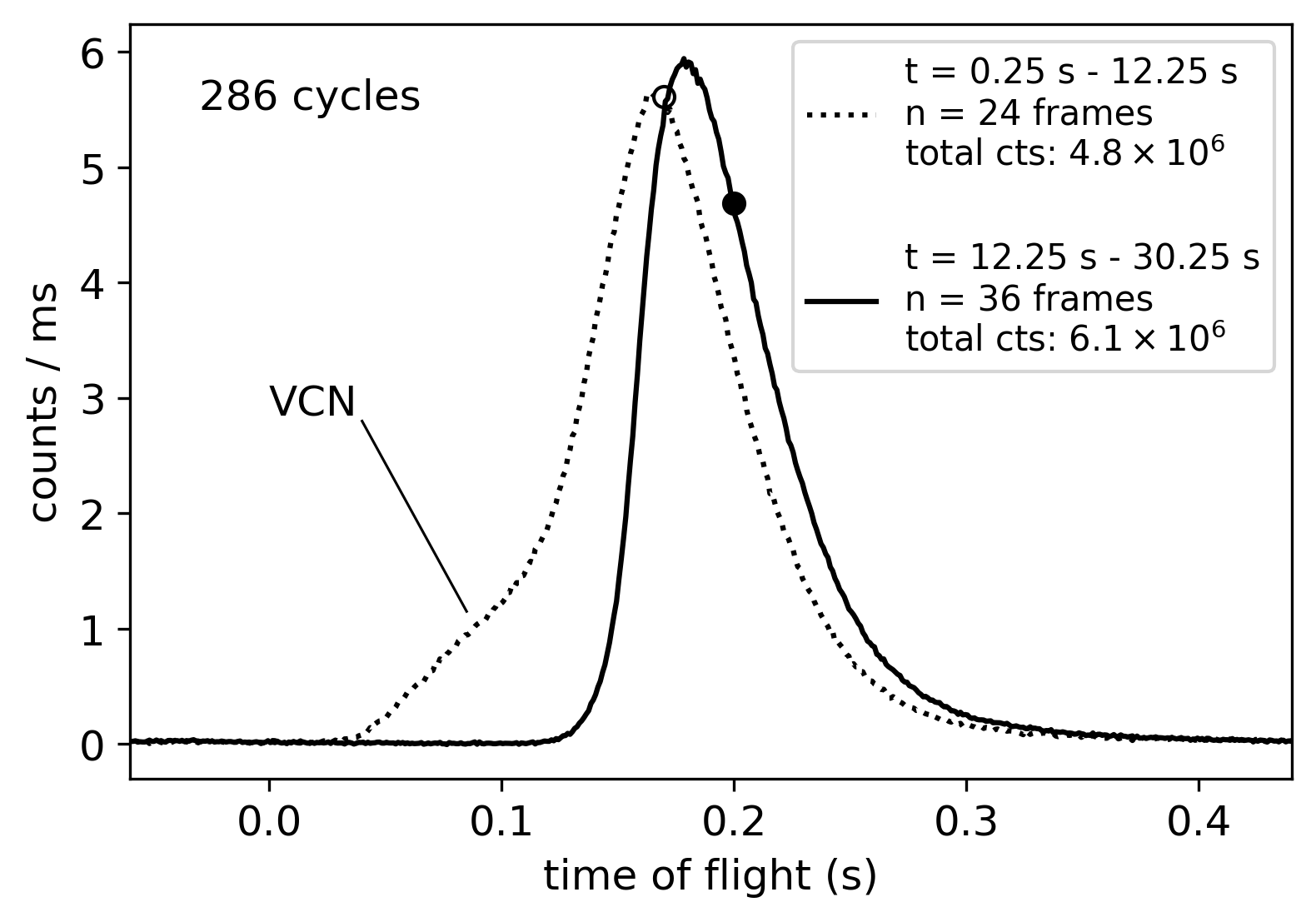}
		}
	\end{center}
	\caption{Averaged TOF spectra per cycle and frame, Eq.\,\eqref{eq:AverageSpectrum}, during the first approximately 12\,\si{\s} (\textbf{dashed line}) and between 12.25\,\si{s} and 30.25\,\si{s} (\textbf{solid line}) after the start of the proton beam pulse. The Poisson errors are smaller than the line width. The \textbf{open} and \textbf{solid markers} indicate the mean $\langle T \rangle$ of the respective distribution. The detection of very cold neutrons (VCN) that are produced during the proton beam pulse is clearly visible by the tail of counts at low time of flight in the averaged spectrum between 0.25\,\si{\s} and 12.25\,\si{\s}. }
	\label{fig:RawTOF}
\end{figure}

The combination of a number of $n$ frames yields the average TOF spectrum
\begin{equation}
\frac{dN}{dt}\Big|_{t_{k}}^{t_{k+n}}(T)=\frac{1}{n}\sum_{j=k}^{k+n}\Big(\frac{dN}{dt}\Big|_{t_j}(T)-R_j\Big)
\label{eq:AverageSpectrum}
\end{equation}
in the corresponding time interval $[t_{k},t_{k+n}]$ after the start of the proton beam pulse. The averaging allows for an easy comparison of spectra and systematic changes between different time intervals, as shown in Fig.\,\ref{fig:RawTOF} for two intervals denoted in the plot.  As a first correction, we subtracted for each frame $k$ individually a rate $R_k$ to reduce the background. The residual background after subtraction was minimized by choosing $R_k$ such that it is the minimum of a 30\,\si{\milli \s} running average of the original TOF spectrum, 
\begin{equation}
R_k = \min_{T}\Bigg\{\frac{1}{30} \sum_{l=0}^{30} \frac{dN}{dt}\Big|_{t_k}(T+l \,  dt)\Bigg\}.
\label{eq:ConstantBackground}
\end{equation}
The running average was used to avoid negative count rates after subtraction. For the first 60 frames ($k\le60$), i.e. the first 30\,\si{\s} after the start of the proton beam pulse, $R_k$ was between 0.04 and 0.5\,\si{\per \milli \s} (on average 0.3\,\si{\per \milli \s}) and lower afterwards.\\

\begin{figure}[t]
	\begin{center}
		\resizebox{0.48\textwidth}{!}{\includegraphics{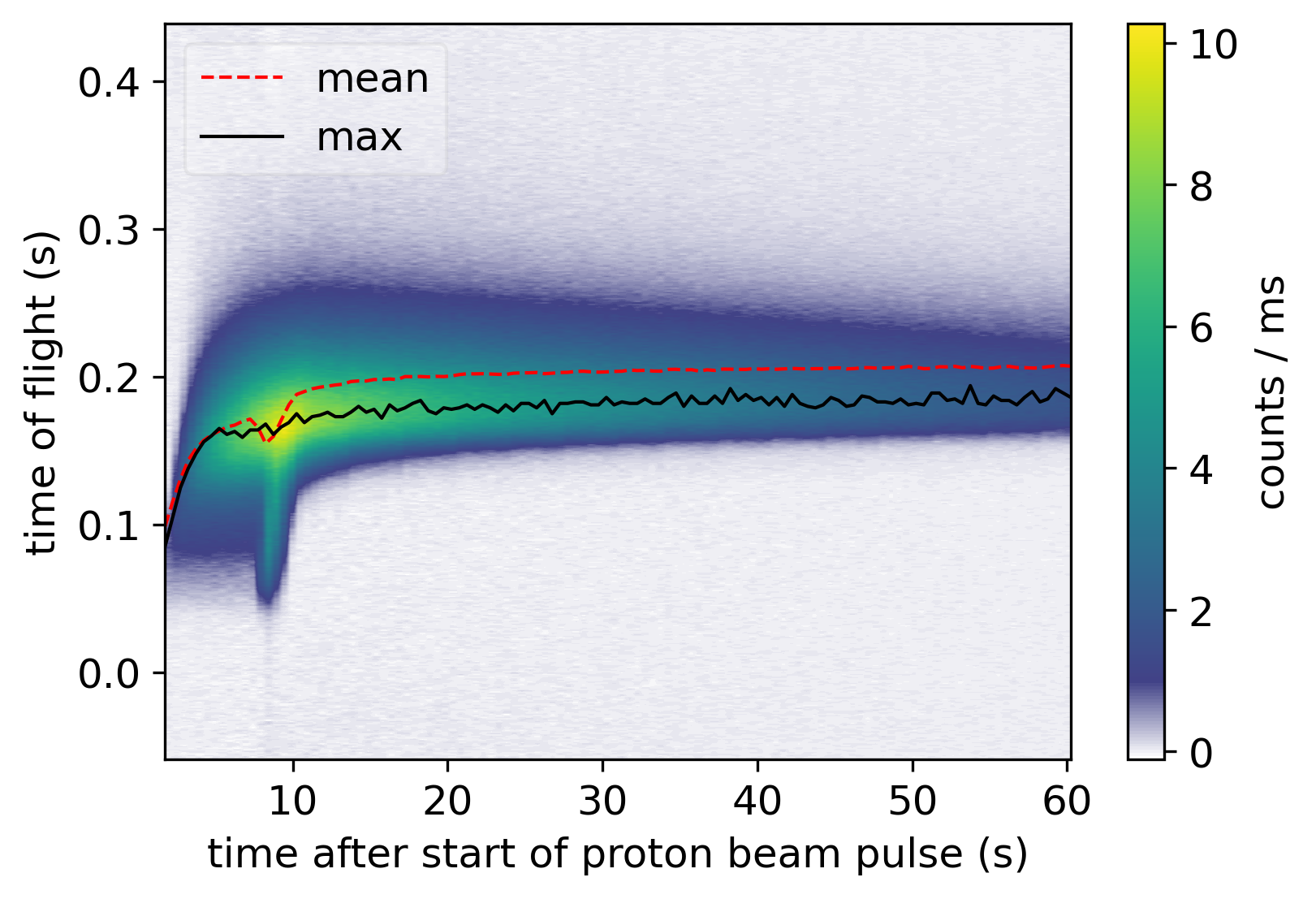}
		}
	\end{center}
	\caption{Evolution of the TOF spectrum during the first 60\,\si{\s} after the start of the 8\,\si{\s} long proton beam pulse. A constant background rate $R_k$, Eq.\,\eqref{eq:ConstantBackground}, was separately determined and removed from the spectrum for each frame $k$ at time $t_k$. The mean time of flight $\langle T \rangle$ (\textbf{red dashed line}) and the position $T\textsubscript{mode}$  of the maximum of the TOF distribution (\textbf{black line}) are indicated in the plot. }
	\label{fig:TOFvsT}
\end{figure}

Figure \ref{fig:TOFvsT} shows the evolution of the TOF spectrum, where the distribution takes its maximum, i.e. the mode $T\textsubscript{mode}$, and the mean time of flight $\langle T \rangle$ during and after the proton beam pulse. The temporary hardening of the spectrum at approximately 8\,\si{\s} was found to be correlated to the closing of the neutron shutter at the bottom of the UCN source storage volume \cite{Rienacker2022}. The hardening is likely due to reflections or scattering of very cold neutrons (VCN) with $v\gtrsim 10\,\si{\m \per \s}$ on the partially closed neutron valve into the West-1 beamline.\\

The further background fitting and subtraction routine follows the method presented in \cite{Daum2014}. In addition to a leakage rate of UCNs through the closed chopper which is approximately constant during a frame (as discussed in section \ref{sec:measurement}), the spectrum also contains UCNs that are reflected non-specularly from surfaces between chopper and detector. After non-specular reflection, the time of flight is no longer a valid measure for the initial velocity parallel to the guide axis. In addition, non-specular reflections and back-reflection of UCNs with velocity components parallel to the guide axis below $v\textsubscript{min}$ can lead to an accumulation of UCNs that are quasi-stored with a short storage time constant $\tau$ in the neutron guide between chopper and detector. Eventually, these quasi-stored UCNs may be deflected towards the detector and counted at times $T>\frac{L}{v\textsubscript{min}} = 0.31\,\si{\s}$, i.e. later than the nominal maximum time of flight . \\

\begin{figure}[b]
	\begin{center}
		\resizebox{0.48\textwidth}{!}{\includegraphics{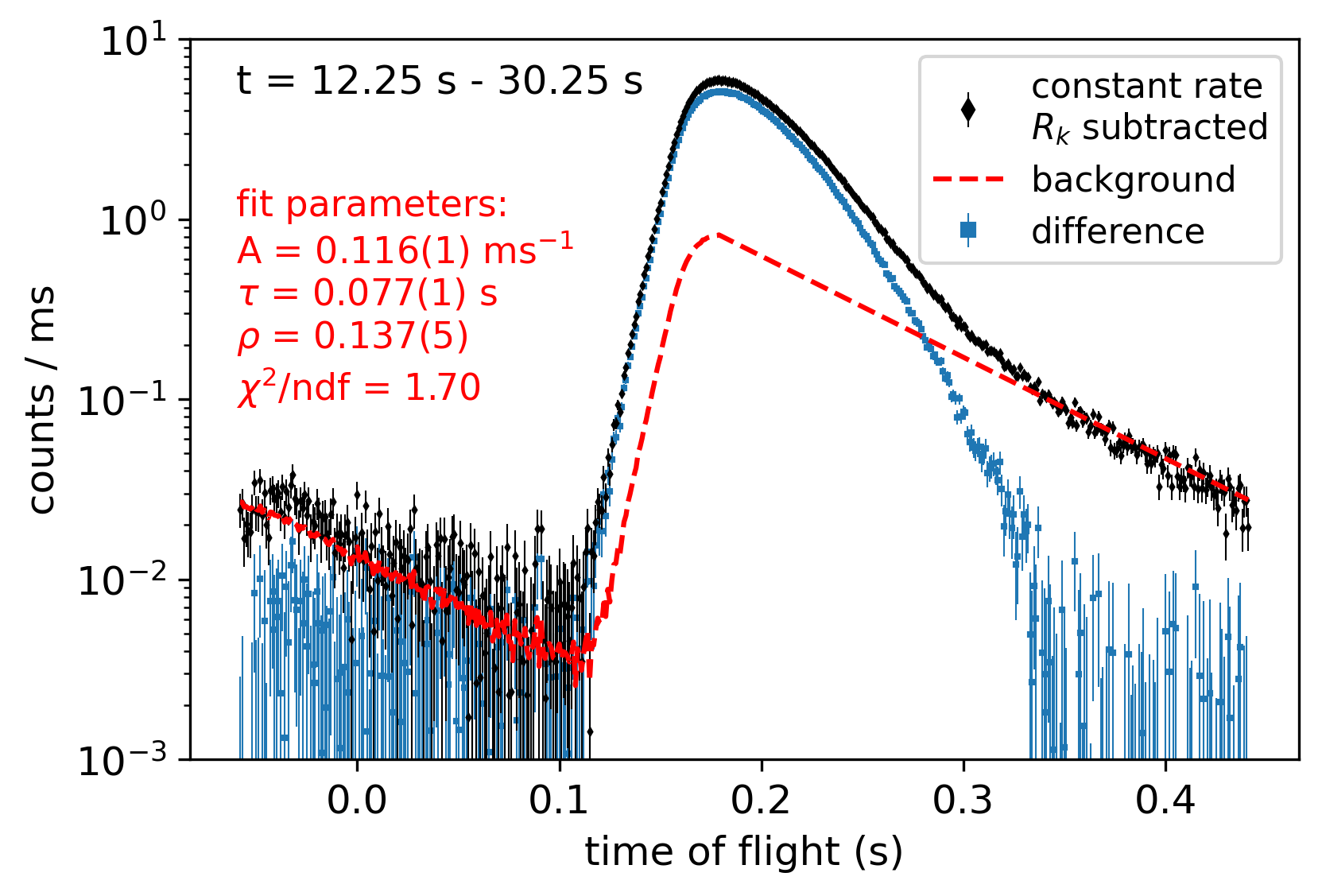}
		}
	\end{center}
	\caption{Plot of the average TOF spectrum between 12.25\,\si{s} and 30.25\,\si{s} after the start of the proton beam pulse, demonstrating the background subtraction method. The \textbf{black diamonds} are the average of the spectra with previously subtracted constant rate $R_k$ per frame k, Eq\,\eqref{eq:ConstantBackground}. The \textbf{red dashed} line indicates the fitted background $B(T)$, Eq.\,\eqref{eq:BackgroundTotal}, according to the procedure explained in the text. The average TOF spectrum with the background removed is the difference (\textbf{blue squares}) of the above.}
	\label{fig:background}
\end{figure}

Figure \ref{fig:background} shows the averaged spectrum between 12.25\,\si{\s} and 30.25\,\si{\s} after subtraction of the rates $R_k$ on a logarithmic scale. We expect a contribution to the residual background due to quasi-stored UCNs in the form of an exponential decay of the count rate after the maximum time of flight of specularly reflected UCNs. Indeed this tail is clearly visible in the plot. Thus, in order to estimate the rate of detected quasi-stored UCNs, we fit the averaged spectrum in Fig.\,\ref{fig:background} at times $T>0.33\,\si{\s}$ with an exponential function
\begin{equation}
b(T) = A \ e^{-(T-0.33\si{\s})/\tau},
\label{eq:Background}
\end{equation}
while simultaneously fitting the same model but shifted by one frame time $b(T+t_f)$ for $T< 0.03\,\si{\s}$, well before the rising edge of the TOF spectrum. Starting the fit interval at $T>0.33\,\si{\s}$, i.e. slightly later than at the nominal maximum time of flight, was required to obtain a low $\chi^2$ for a single exponential fit. We confirmed that including a constant offset parameter $R$ in the fit to the average TOF spectra, without previously removing a constant rate $R_k$ for each frame individually, leads to a similar result.  \\

We assume that at the beginning of the frame for $T<T\textsubscript{mode}$, i.e. before the maximum of the time-of-flight distribution, the detected rate of non-specularly reflected UCNs follows the same time distribution as the specularly reflected UCNs. The systematic implications of this assumption are discussed below. Thus, the function $b(T)$ with best fit parameters $(A,\tau)$ is extrapolated to the region $T\ge T\textsubscript{mode}$. For $T<T\textsubscript{mode}$ we add a fraction $\rho$ of the TOF spectrum to the contribution $b(T+t_f)$ of quasi-stored UCNs from the previous frame to obtain a smooth transition of the background function $B(T)$ (Fig.\,\ref{fig:background} dashed red line) before and after the maximum of the TOF spectrum. In general, for the averaged TOF spectrum in the interval $[t_{k},t_{k+n}]$ the background function is thus given by
\begin{equation}
\begin{aligned}
&B|_{t_{k}}^{t_{k+n}}(T) = \\
&=\begin{cases}
(1-\rho) \, b(T+t_f) + \rho \, \frac{dN}{dt}|_{t_{k}}^{t_{k+n}}(T) &; T<T\textsubscript{mode}\\
b(T) &; T\ge T\textsubscript{mode}
\end{cases}
\label{eq:BackgroundTotal}
\end{aligned}
\end{equation}
with
\begin{equation}
\rho = \frac{b(T\textsubscript{mode})-b(T\textsubscript{mode}+t_f)}{\frac{dN}{dt}|_{t_{k}}^{t_{k+n}}(T\textsubscript{mode})-b(T\textsubscript{mode}+t_f)}
\label{eq:BackgroundScale}.
\end{equation} 
Note that for better readability we have omitted the notation to designate the time interval $[t_{k},t_{k+n}]$ on the exponential function $b$ and the parameters $(A, \tau, \rho)$. However they are determined from the spectrum averaged over the chosen range and thus depend on the corresponding time interval.
\\

\begin{figure}[t]
	\begin{center}
		\resizebox{0.48\textwidth}{!}{\includegraphics{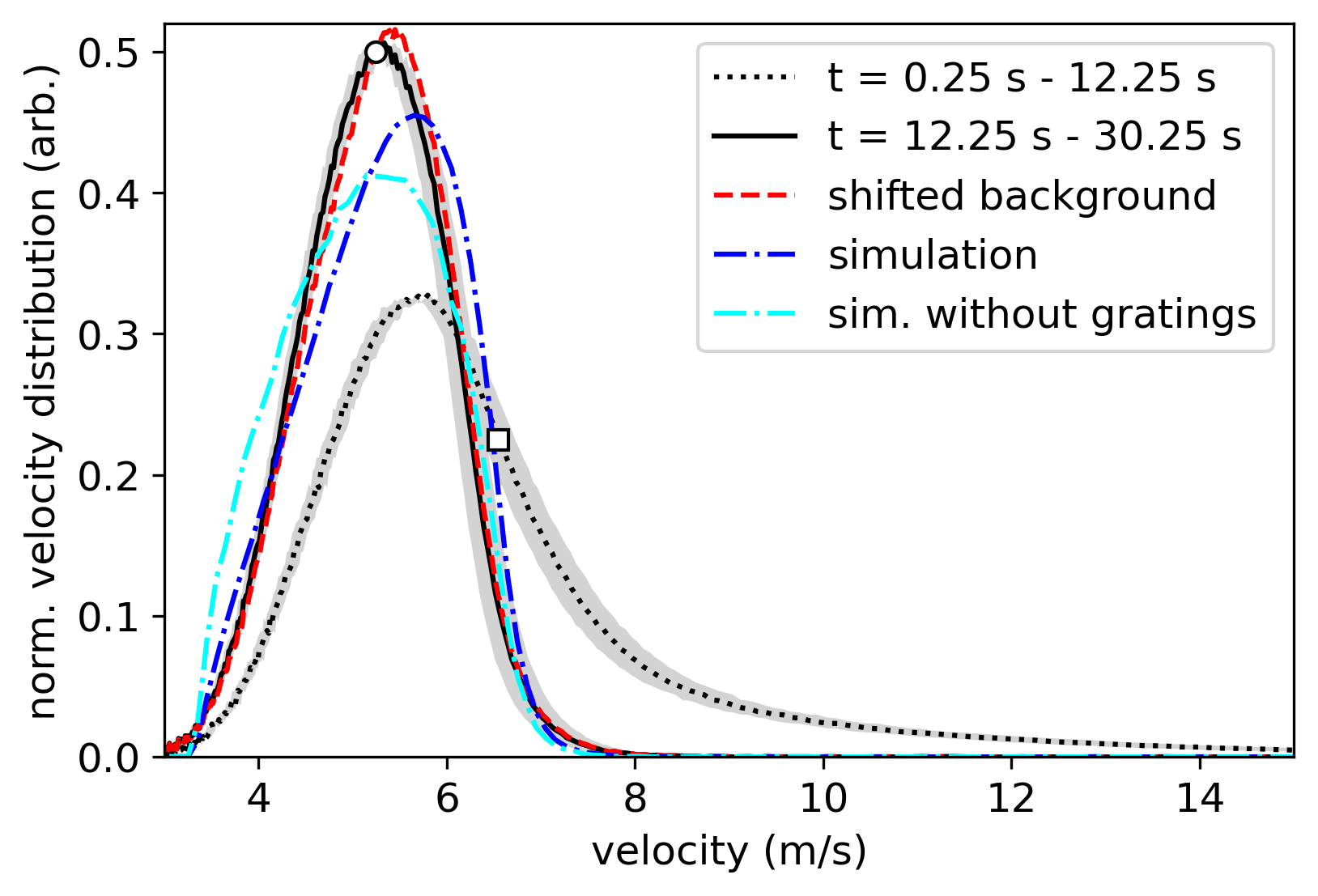}
		}
	\end{center}
	\caption{Normalized velocity spectra during the first approximately 12\,\si{\s} (\textbf{dashed line}) and between 12.25\,\si{s} and 30.25\,\si{s} (\textbf{solid line}) after the start of the proton beam pulse. The shaded regions mark the uncertainty on the velocity bins due to timing uncertainties of the chopper. The statistical uncertainty is smaller than the line width. The markers indicate the mean of the velocity distributions of 6.5(1)\,\si{\m \per \s} and 5.2(1)\,\si{\m \per \s} in the corresponding time intervals. The \textbf{red dashed line} demonstrates the effect on the resulting velocity spectrum between 12.25\,\si{s} and 30.25\,\si{s} if the maximum of the background curve (red dashed line in Fig.\,\ref{fig:background}) is shifted by $+30\,\si{\milli \s}$ as discussed at the end of section \ref{sec:analysis}.  The \textbf{cyan dot-dashed line} shows the axial velocity spectrum at the detector position (between 12\,\si{s} and 30\,\si{s} after the proton beam pulse) obtained from a Monte Carlo simulation and model of the UCN source presented in \cite{Bison2020} and \cite{Bison2022}. The \textbf{blue dot-dashed line} is the corresponding result if the chopper gratings are implemented in the simulation to correctly consider the collimation effect outlined in section \ref{sec:measurement}.
	}
	\label{fig:velocity}
\end{figure}

For the time range depicted in Fig.\,\ref{fig:background}, the ratio of the integrated background, including the offsets $R_k$, to the integral of the final TOF spectrum (Fig.\,\ref{fig:background}, blue markers) is 13\,\%. We checked the impact of our assumption about the time distribution of non-specularly reflected UCNs on the final velocity spectrum by shifting the mode of the background curve (Fig.\,\ref{fig:background}, red dashed line) by 30\,\si{\milli \s}, half of the width of the distribution. As a consequence, the mean of the resulting velocity distribution was found to be shifted by at most 0.05\,\si{\m \per \s} (which we included in our uncertainty in Fig.\,\ref{fig:meanvel}) and the mode by up to 0.3\,\si{\m \per \s}, see Fig.\,\ref{fig:velocity}. Similarly, an extreme change of the background model in the signal region to a linear interpolation from $T=0.1\,\si{\s}$ to $T=0.33\,\si{\s}$ was found to shift the mean velocity by less than 0.06\,\si{\m \per \s}.

\section{Results and discussion}
\label{sec:results}

Figure \ref{fig:velocity} shows the final results of the axial velocity spectra
\begin{equation}
\frac{dN}{dv}\Big|_{t_{k}}^{t_{k+n}}(v) = \frac{L}{v^2} \Big( \frac{dN}{dt}\Big|_{t_{k}}^{t_{k+n}}(L/v)-B|_{t_{k}}^{t_{k+n}}(L/v)\Big)
\end{equation}
for two time intervals after the proton beam pulse. The background $B$ was subtracted from the TOF spectra as described in section \ref{sec:analysis}. Due to the short opening time and commonly observed numerical instabilities in deconvolving measured data, no deconvolution of the TOF spectra with the chopper resolution function was applied. Instead we have considered the chopper opening time as a timing uncertainty as mentioned in the previous section and propagated the corresponding effect to the velocity bins for the following analysis.\\

\begin{figure}[b]
	\begin{center}
		\resizebox{0.48\textwidth}{!}{\includegraphics{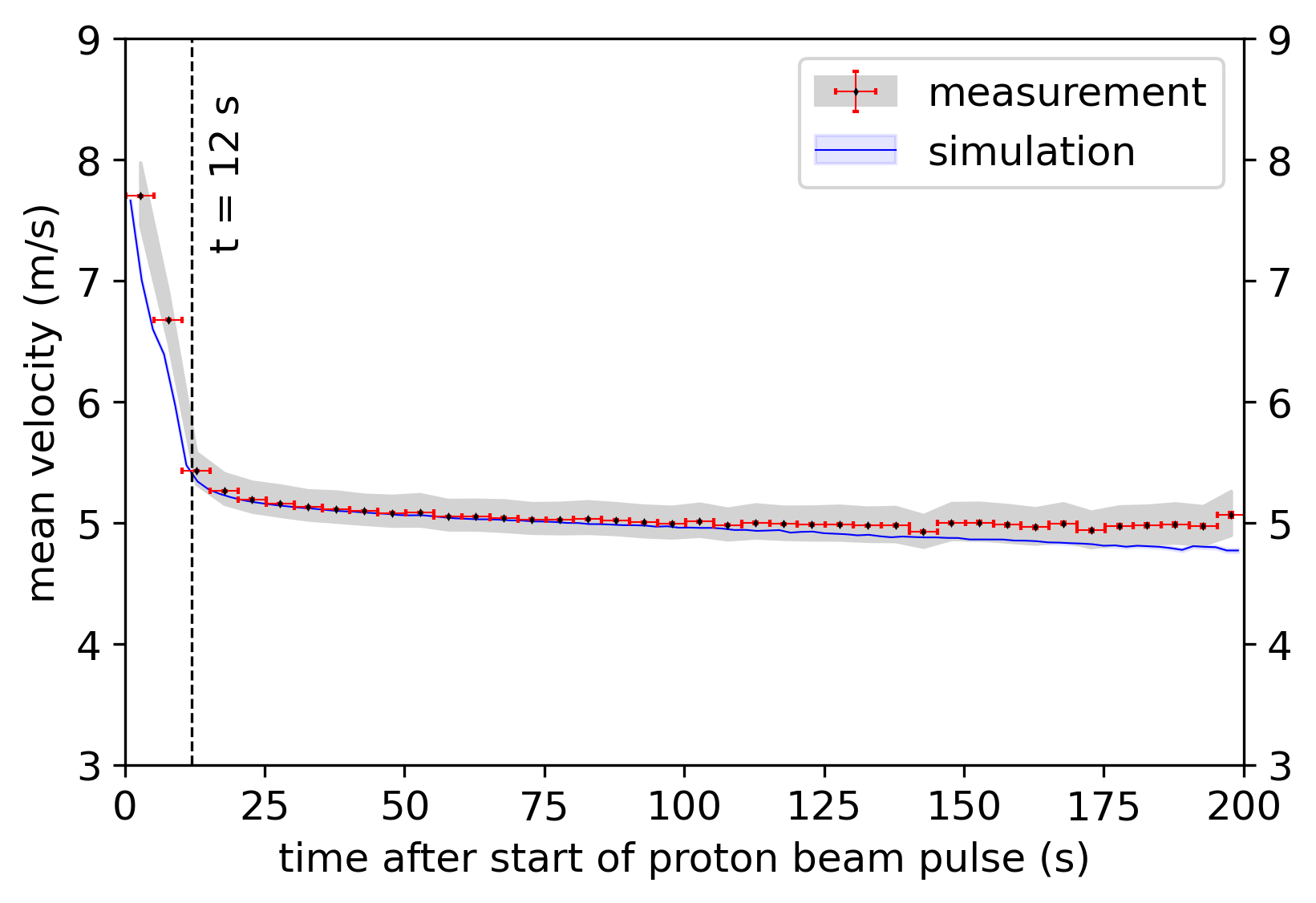}
		}
	\end{center}
	\caption{Evolution of the measured mean velocity parallel to the guide axis during 200\,\si{\s} after the start of the 8\,\si{\s} proton beam pulse. The \textbf{red error bars} indicate the 5\,\si{\s} binning and the statistical error, while the \textbf{grey band} represents the effect of the chopper timing uncertainties. For the first 12\,\si{\s}, i.e. during and shortly after the pulse, the mean velocity is significantly higher than at later times, where the mean velocity slowly reduces from about 5.4\,\si{\m \per \s} to 5.0\,\si{\m \per \s}. The \textbf{blue line} and \textbf{shaded region} denote the mean and statistical uncertainty of the velocity component parallel to the guide axis at the detector position, obtained from a Monte Carlo simulation with a model of the source described and calibrated in \cite{Bison2020,Bison2022} including the chopper gratings. 
	}
	\label{fig:meanvel}
\end{figure}

The evolution of the mean velocity
parallel to the guide axis in 5\,\si{\s} intervals during 200\,\si{\s} after the start of the proton beam pulse, is shown in Fig.\,\ref{fig:meanvel}. The mean velocity of the neutrons reduces from approximately 7.7(1)\,\si{\m \per \s} to 5.4(1)\,\si{\m \per \s} within the first 12\,\si{\s} after the start of the proton beam pulse. The initial softening of the spectrum during the proton beam pulse correlates with the increase of UCN density in the source storage volume. This leads to a dominance of UCNs over non-storable VCN produced at an approximately constant rate. After approximately 12\,\si{\s} the spectrum consists mostly of UCNs (see Fig.\,\ref{fig:velocity}) and the mean velocity decreases slowly to 5.1(1)\,\si{\m \per \s} at 30.25\,\si{\s}, comparable to the typical filling times of storage experiments. Indeed, a gradual softening of the spectrum during the storage of UCNs is expected due to velocity dependent losses in the source storage volume and neutron guides \cite{Bison2020}.\\

We compared our results to the UCN velocity component parallel to the beamline axis obtained from a MCUCN \cite{Zsigmond2018} Monte Carlo simulation. We used the simulation model of the source described and calibrated in \cite{Bison2020} and \cite{Bison2022} and adjusted the geometry to include the stainless steel flanges and guide, as well as the 1\,\si{\m} long glass guide of the chopper setup. The UCN velocity component parallel to the guide axis was tallied at the detector position, behind the AlMg3 entrance window. We have implemented the capture and conversion efficiency of the 200 nm thick ${}^{10}$B layer of the Cascade detector following \cite{Jenke2013}, which distorts the spectrum below 7\,\si{\m \per \s} by less than 3\,\%. The surface parameters of the guides were set to those found in \cite{Bison2022} for the UCN source beamlines.\\

The result from the simulation model as described above is shown by the cyan dot-dashed curve in Fig.\,\ref{fig:velocity}. As mentioned in section \ref{sec:measurement}, the excess below approximately 5\,\si{\m \per \s} of the simulation over the measured spectrum can be explained by the collimation effect of the chopper. To address this effect we have included an implementation of the chopper gratings with dimensions as described in section \ref{sec:measurement} in our full simulation model. The titanium gratings were modelled to be fully absorbing and placed between the stainless steel and coated glass guide in the simulation model. The displacement of the two gratings along the motor axis was varied to achieve different openings. We sampled the opening function of the chopper at 11 equidistant times with slit openings of 0.4, 1.1, 1.9, 2.5, 2.9, 3, 2.9, 2.5, 1.9, 1.1 and 0.4\,\si{\milli \m} and directly summed the normalized velocity spectra to obtain the blue dot-dashed line in Fig.\,\ref{fig:velocity}. Even though there are some discrepancies between the shape of the measured and simulated spectra, the evolution of the mean velocity obtained from this simulation shown in Fig.\,\ref{fig:meanvel} reproduces the observed softening of the velocity spectrum during and after the proton beam pulse well. This indicates that the velocity dependent loss mechanisms in the source storage volume and neutron guides are modelled correctly to a large degree by the simulation. \\

\begin{figure}[t]
	\begin{center}
		\resizebox{0.48\textwidth}{!}{\includegraphics{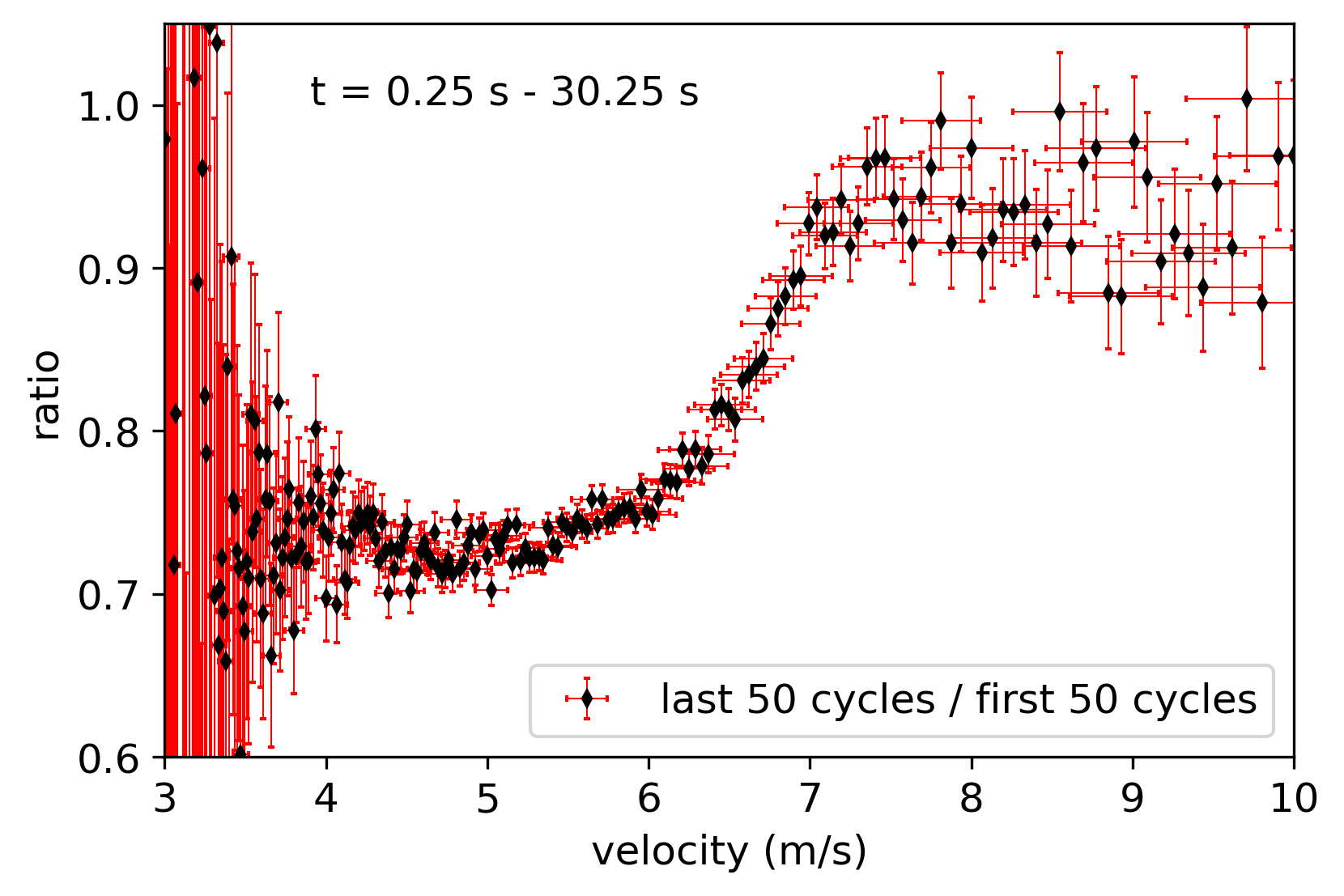}
		}
	\end{center}
	\caption{Ratio of the average velocity spectra between 0.25 and 30.25\,\si{\s} after the proton beam pulse of the last 50 measurement cycles (cycle 237 - 286) and the first 50 cycles (cycle 1 - 50).}
	\label{fig:ratio}
\end{figure}

In reference \cite{Anghel2018} it was reported that heat deposition by radiation from the spallation target can cause a built-up of solid deuterium (sD$_2$) frost on the surface of the sD$_2$ moderator over time. Simulations showed that frost layers cause a hardening of the UCN energy spectrum (Fig.\,20 of \cite{Anghel2018}) due to velocity-dependent back-scattering of UCNs from the optical potential of isotropically distributed sD$_2$ disks. A conditioning procedure was developed at the PSI UCN source to restore the surface quality. We studied the evolution of the average velocity spectrum on long time scales by performing our analysis on subsets of measurement cycles. Figure \ref{fig:ratio} shows the ratio of the velocity spectrum between 3 and 10\,\si{\m \per \s} obtained from the last 50 cycles, and the first 50 cycles, recorded during roughly one day (i.e. 286 cycles). A conditioning procedure was applied shortly before the first cycle of our measurements. While the mean velocity of the first 50 cycles and the last 50 cycles differs only by about 0.1\,\si{\m \per \s}, the ratio shown in Fig.\,\ref{fig:ratio} reduces quickly from close to one for velocities above approximately 7.2\,\si{\m \per \s} to about 75\,\% for lower velocities. This indicates a hardening of the spectrum by additional losses of slower UCN, consistent with the observed effects of surface degradation of the solid deuterium moderator due to the built-up of sD$_2$ frost layers \cite{Anghel2018}. 

\section{Conclusion}

The UCN velocity spectrum parallel to the guide axis at beamport West-1 was measured by time-of-flight spectroscopy using a neutron chopper. A systematic background subtraction was necessary due to contributions of a constant leakage of UCNs through the closed chopper and non-specular reflections on surfaces between chopper and detector. Furthermore the collimation effect of the chopper gratings must be considered for a correct comparison to the UCN source simulation model. We measured the decrease of the mean UCN velocities during and after the proton beam pulse from about 7.7\,\si{\m \per \s} to 5\,\si{\m \per \s}. We have also determined the evolution of the UCN energy spectrum during source operation, where sD$_2$ frost build-up on the solid deuterium surface causes additional losses, predominantly of slower UCN.\\

The measurement of the velocity spectra at beamport West-1 is another step towards the full neutronics characterization \cite{Bison2022a,Bison2020,Becker2015,Bison2022} of the PSI UCN source. Our results show good overall agreement with simulations. Together with an accurate understanding on how the chopper affects the measured axial velocity spectrum, our results can be used for further refinements of the UCN source simulation model and in the study of possible improvements of the UCN source.

\section*{Acknowledgments}

We would like to thank the referees to point us towards the significant collimation effect of the chopper and T. Jenke for useful discussion about this effect. We acknowledge the PSI proton accelerator operations section, all colleagues who have been contributing to the UCN source operation at PSI, and especially the BSQ group which has been operating the PSI UCN source during the measurements, namely B. Blau, P. Erisman and also S. Grünberger. Excellent technical support by M. Meier, L. Noorda and M. Schaufelbühl is acknowledged.  This work was supported by the Swiss National Science Foundation Projects 
163413, 
169596, 
172626, 
178951, 
188700 
and 200441. 

\bibliographystyle{spphys}  

\end{document}